\documentclass[a4paper,11pt]{article}
\pdfoutput=1 

\usepackage{jcappub} 

\usepackage[T1]{fontenc} 
\usepackage{subfig}

\def\be{\begin{equation}}
\def\ee{\end{equation}}
\def\barr{\begin{array}}
\def\earr{\end{array}}
\def\bea{\begin{eqnarray}}
\def\eea{\end{eqnarray}}
\def\bfig{\begin{figure}}
\def\efig{\end{figure}}

\newcommand{\nn}{\nonumber}

\title{Effects of QCD Equation of State on the Stochastic Gravitational Wave Background}

\author[a]{Sampurn Anand,}
\author[b]{Ujjal Kumar Dey,}
\author[a]{Subhendra Mohanty}

\affiliation[a]{Physical Research Laboratory,\\ Ahmedabad 380009, India}
\affiliation[b]{Centre for Theoretical Studies, Indian Institute of Technology Kharagpur,\\ Kharagpur 721302, India}

\emailAdd{sampurn@prl.res.in}
\emailAdd{ujjal@cts.iitkgp.ernet.in}
\emailAdd{mohanty@prl.res.in}

\abstract{Cosmological phase transitions can be a source of Stochastic Gravitational Wave (SGW) background. Apart from the dynamics of the phase transition, the characteristic frequency and the fractional energy density $\Omega_{\rm gw}$ of the SGW depends upon the temperature of the transition. In this article, we compute the SGW spectrum in the light of QCD equation of state provided by the lattice results. We find that the inclusion of trace anomaly from lattice QCD, enhances the SGW signal generated during QCD phase transition by $\sim 50\%$ and the peak frequency of the QCD era SGW are shifted higher by $\sim 25\%$ as compared to the earlier estimates without trace anomaly. This result is extremely significant for testing the phase transition dynamics near QCD epoch.}


\begin{document}
\maketitle
\flushbottom

\section{Introduction}
\label{sec:intro}

A new era of astronomy and cosmology was ushered in  by the discovery of gravitational waves (GW) from merging black holes by the LIGO collaboration \cite{Abbott:2016blz}. LIGO detectors operate in the high frequency range ($10 -10^3$) Hz to detect sources like compact binary inspirals. In order to detect sources like massive binaries, Supernovae etc., which produces low frequency GW signal, space-based detectors such as eLISA \cite{Klein:2015hvg}, expected to operate between $(10^{-5} -1)$ Hz, have been proposed. Furthermore, experiments like Pulsar Timing Array (PTA) \cite{IPTA-2013} and Square Kilometer Array (SKA) \cite{Dewdney-ska,Kramer:2004hd} can measure frequency range as low as $10^{-9}$ Hz and promise to explore the stochastic gravitational wave backgrounds. Therefore, these upcoming detectors will begin the multi-wavelength GW astronomy as a means of understanding the universe in much greater detail.
On the theoretical front, modeling sources of GW signals is important. In particular, the stochastic gravitational wave background which encapsulates the information about the early universe is of great interest. It has been argued that if the cosmological phase transitions are first order and lasts for sufficiently long duration then they can be a potential source of very low frequency stochastic gravitational wave background \cite{Witten:1984rs, Hogan:1984hx, 1986MNRAS.218..629H, Turner:1990rc}. Within the purview of the standard model of particle physics, at least two transitions, namely electroweak phase transition at $T_* \sim 100$ GeV and QCD phase transition at $T_*\sim 0.1$ GeV, have taken place. The former transition is associated with electroweak symmetry breaking and latter with the breaking of chiral symmetry. Although, it has been  argued that the above mentioned transition proceed through a smooth crossover \cite{Aoki:2006we, Bhattacharya:2014ara, Kajantie:1995kf, Kajantie:1996mn}, there are models beyond standard model which can generate strong first order transition at electroweak \cite{Grojean:2006bp, Delaunay:2007wb, Huber:2007vva, Dev:2016feu, Balazs:2016tbi} and QCD \cite{Schwarz:2009ii,Schwaller:2015tja} scale.
Recently, QCD lattice results \cite{DeTar:2009ef} have shown that in the presence of strong interactions, the pressure $p$ deviates from its value in the radiation dominated epoch {\it i.e.}, $1/3$ of the energy density $\rho$. This interaction measure is characterized in terms of trace anomaly and it can have interesting cosmological consequences. For instance, the prediction of Weakly Interacting Massive Particles (WIMPs) relic density differs \cite{Drees:2015exa} from the phenomenological models or pure glue lattice QCD calculations. Therefore, it would be interesting to explore the consequences of trace anomaly on the SGW signal generated during QCD phase transition where the anomaly is prominent.

In this article, we revisit the energy spectrum calculation of GW in the light of QCD equation of state provided by recent lattice results. GW once produced, traverse through the space-time almost undisturbed till today. Consequently, they are unique probe of the phase transitions dynamics and hence, it is worth exploring. This paper is organized as follows: in section \ref{s:calc} we include the QCD equation of state and calculate the general expression of GW spectrum. In section \ref{s:sourcegw} will discuss about the processes during first order phase transition by which GW can be generated. Finally, we discuss the results in section \ref{s:result} and conclude in \ref{s:concl}.

\section{Gravitational wave spectrum revisited} 
\label{s:calc}
In order to estimate the observable gravitational wave background today, we propagate the GW from the epoch of phase transition, where they are generated, to the current epoch. In order to do so, we consider the Boltzmann equation which takes simple form, $\frac{d}{dt}(\rho_{gw}\, a^4) = 0\, ,$ due to the fact that the gravitational waves are essentially decoupled from rest of the universe. Assuming that the universe has expanded adiabatically since the phase transition, which implies that the entropy per comoving volume $S \propto a^3\, g_s\, T^3 $ remains constant, i.e., $\dot S/S=0\,$, gives us the time variation of temperature as  
\be
\frac{dT}{dt} = -H T\,\left(1+ \frac{T}{3g_s}\frac{dg_s}{dT} \right)^{-1}\,
\label{Tdot}
\ee
where $g_s$ is the effective number of relativistic degrees of freedom that contributes to the entropy density. Using the relation given in Eq. (\ref{Tdot}), it can be shown that
\be
\frac{a_*}{a_0} = \exp \bigg[\int_{T_*}^{T_0} \frac{1}{T}\big(1+ \frac{T}{3g_s}\frac{dg_s}{dT}\big)\, dT\bigg]
\ee
and the energy density of the gravitational waves at current epoch is given as
\be
\rho_{\rm gw}(T_0) = \rho_{\rm gw}(T_*)\,\exp\bigg[\int_{T_*}^{T_0}
  \frac{4}{T}\left(1+ \frac{T}{3g_s}\frac{dg_s}{dT}\right)\, dT\bigg]
\label{eq:rho_gw}
\ee
where `` * '' denotes the quantities at the epoch of phase transition and `` 0 '' denotes the same at current epoch (and this convention has been followed in the rest of the paper).
\begin{figure}[!htbp]
\begin{center}
\subfloat[\label{fig:pbyrho}]{
\includegraphics[width=3.2in,height=2.6in,angle=0]{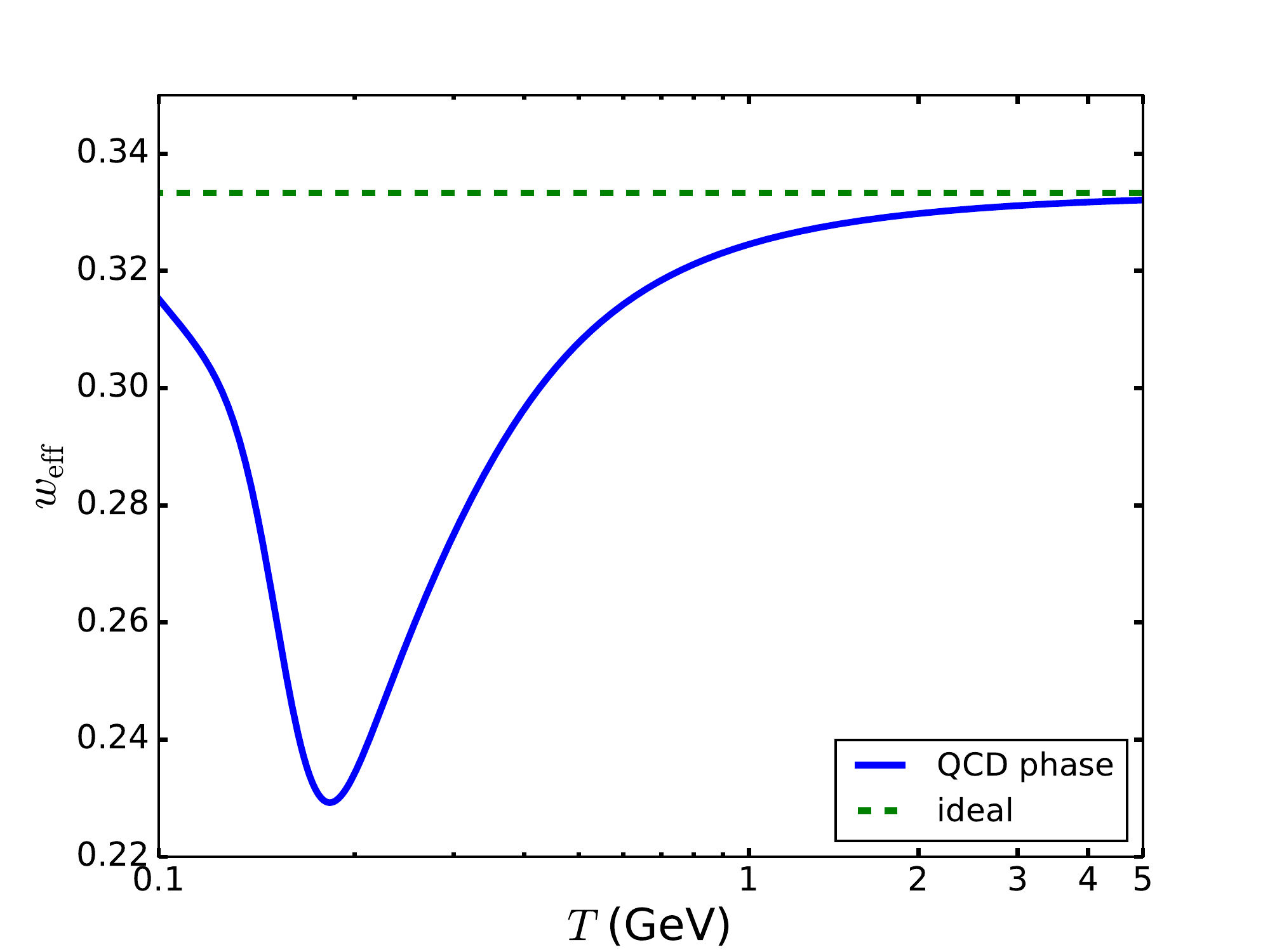}}~~~~
\hspace{-0.6cm}
\subfloat[\label{fig:ast-a0}]{
\includegraphics[width=3.2in,height=2.6in,angle=0]{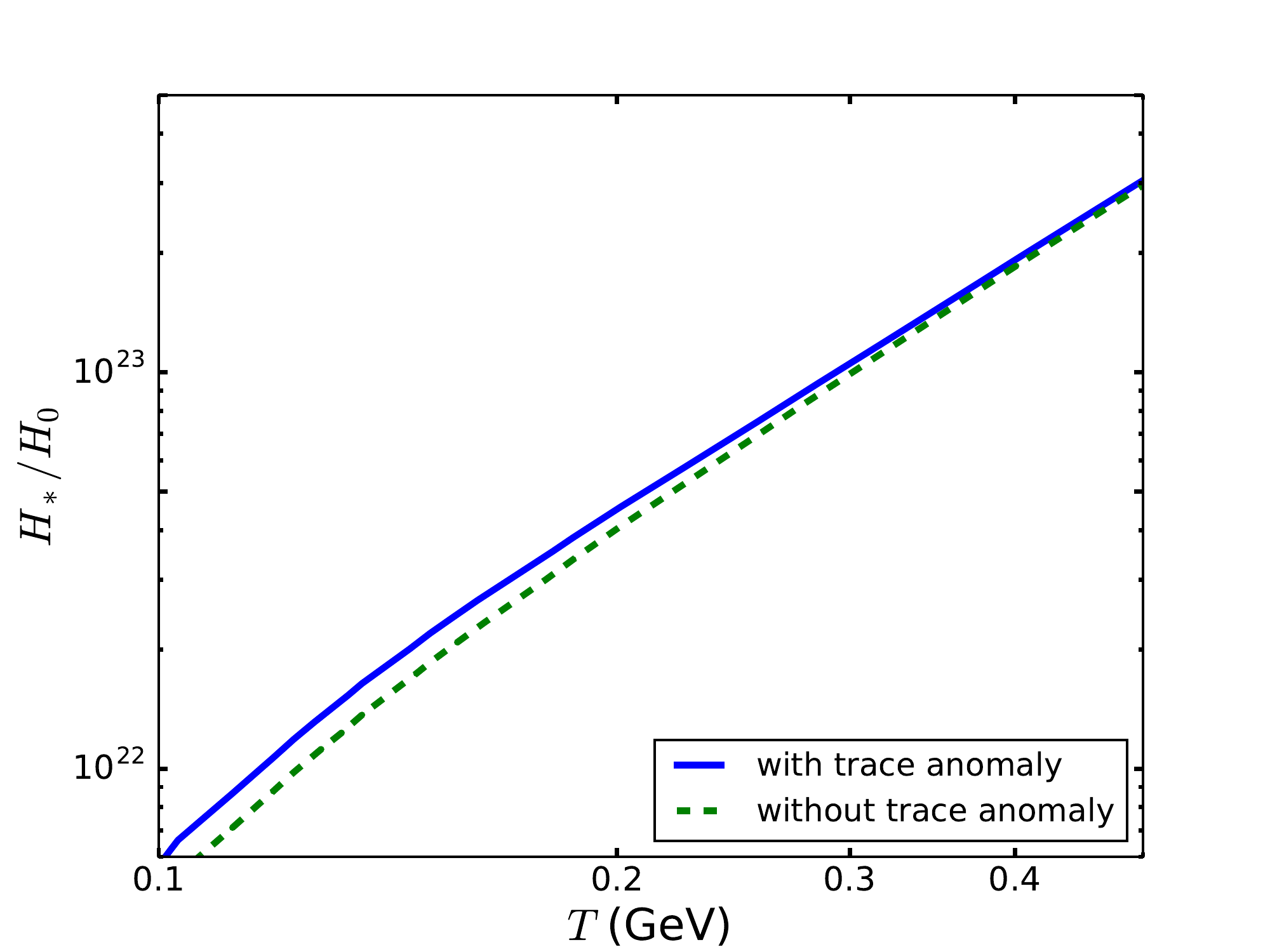}}
\caption{(a) Variation of equation of state with temperature.
   (b) $H_*/H_0$ as a function of transition temperature. In both the figures, dashed line represents the ideal gas equation of state where $p = \rho/3$ while solid line represents the equation of state as parametrized in Ref.\cite{Bazavov:2014pvz}.~~~~~~~~~~~~}
\end{center}
\end{figure}
With this, we can define the  density parameter of the gravitational waves at current epoch, $\Omega_{\rm gw} = \frac{\rho_{\rm gw}(T_0)}{\rho_{\rm cr}(T_0)}$, as
\be
\Omega_{\rm gw}=\Omega_{\rm gw*}\bigg(\frac{H_*}{H_0}\bigg)^2
\exp\bigg[\int_{T_*}^{T_0}
  \frac{4}{T}\left(1+\frac{T}{3g_s}\frac{dg_s}{dT}\right)\, dT\bigg]
\label{eq:gw_den}
\ee
where $\frac{\rho_{\rm cr}(T_*)}{\rho_{\rm cr}(T_0)} =\big(\frac{H_*}{H_0}\big)^2 $ has been used.
To evaluate the ratio of the Hubble parameter at the epoch of transition to that of its value today, we consider the continuity equation, given as: $\dot\rho_{t} = -3H\rho_{t} \big(1+ p_{t}/\rho_{t}\big)$, with $\rho_t (p_t)$ being the total energy (pressure) density of the universe and dot denotes the derivative with respect to cosmic time. Using Eq. (\ref{Tdot}) we can express the continuity equation in terms of temperature as,
\be
\frac{d\rho_t}{\rho_t} = \frac{3}{T}\,\left(1+ w_{\rm eff}\right)
   \bigg(1+ \frac{T}{3g_s}\frac{dg_s}{dT}\bigg)\, dT\, ,
   \ee
where $w_{\rm eff} = p_t/\rho_t$ is the effective equation of state parameter. Thus, energy density at the time of phase transition, obtained by integrating the above equation from some early time in the radiation dominated epoch where temperature is $T_r$ till the epoch of phase transition, is given as,
\be
\rho(T_*) = \rho(T_r)\, \exp\bigg[\int_{T_r}^{T_{*}}
  \frac{3}{T}\,(1+ w_{\rm eff} )
  \left(1+ \frac{T}{3g_s}\frac{dg_s}{dT}\right)\, dT\bigg].
\label{eq:energy}
\ee
For ultra-relativistic, non-interacting particles $w_{\rm eff} =  1/3$. However, around $T\sim$ a few hundred MeV, the strong interaction leads to a deviation from $w_{\rm eff} =1/3$. The equation of state around QCD epoch can be computed using the parametrization of the pressure due to $u, d, s$ quarks and gluons, as given in Ref.\cite{Bazavov:2014pvz},
\be
\frac{2p}{T^4} =\big(1 + {\rm tanh[c_\tau(\tau - \tau_0)]}\big)
\bigg(
\frac{p_i + a_n/\tau + b_n/\tau^2 + d_n/\tau^4}{ 1+ a_d/\tau + b_d/\tau^2 + d_d/\tau^4}\bigg)
\label{eq:pressure}
\ee
where $\tau = T/T_c$ with $T_c = 154$ MeV being the QCD transition temperature and $p_i = (19\,\pi^2)/36$ is the ideal gas value of $p/T^4$ for QCD with three massless quarks. The value of numerical coefficients in Eq.~(\ref{eq:pressure}) are : $c_\tau = 3.8706$, $a_n = -8.7704$, $b_n = 3.9200$, $d_n = 0.3419$, $a_d = -1.2600$, $b_d =0.8425$, $d_d = -0.0475$ and $\tau_0 = 0.9761$. Using the expression for pressure given in Eq.~(\ref{eq:pressure}), the energy density can be computed from the trace anomaly \cite{Cheng:2007jq} given as,
\be
\frac{\rho - 3p}{T^4} = T\,\frac{d}{dT}\bigg(\frac{p}{T^4}\bigg)\,.
\ee
Note that this parametrization automatically approaches the ideal gas value for $T\gg T_c$ because of the ${\rm tanh}$ function (see Fig.~(\ref{fig:pbyrho})). It has also been shown in Ref.\cite{Bazavov:2014pvz} that Eq.~(\ref{eq:pressure}) matches with perturbative calculations at higher temperature. Therefore, it can be used to describe the contribution form $u, d, s$ quarks and gluons for all temperatures above 100 MeV. At temperatures lower than $T_c$, the thermodynamic behavior of QCD is well described by hadron resonance gas model. However, for the present calculation this is not required as we are interested in the gravitational waves from phase transition above QCD scale.
Apart from quarks and gluons, contribution to the total energy density and pressure will come from other particles as well. However, energy density and pressure of a non-relativistic particle is exponentially smaller than that of the relativistic particles. Consequently, we include the energy density $\,\rho_{\rm rel}$ and pressure $\,p_{\rm rel}$ of relativistic particles, which is given as: $\rho_{\rm rel} = (\pi^2/30)\big(\sum_{i={\rm bosons}} g_i  + \sum_{j={\rm fermions}}(7/8) g_j\big)T^4$ and  $p_{\rm rel} = \rho_{\rm rel}/3$, respectively. Including the contribution of relativistic particles along with quarks and gluons, the  effective equation of state is shown in Fig.~(\ref{fig:pbyrho}). Note that the trace anomaly vanishes at $T\sim$ 4 GeV indicating the radiation dominated epoch. However, trace anomaly is present even at $T\sim$ 2 GeV and decreases the value of $w_{\rm eff}$ from 1/3.

With the above inputs, we can numerically integrate Eq.~(\ref{eq:energy}) and use the relation $H_*^2 = \rho_*/(3\,m_p^2)$ to define
\begin{align}
 \bigg(\frac{H_*}{H_0}\bigg)^2 = 
 \Omega_{r0} \bigg(\frac{a_0}{a_r}\bigg)^4  \exp\bigg[\int_{T_{r}}^{T_{*}}
    \frac{3(1 + w_{\rm eff})}{T}\left(1+ \frac{T}{3g_s}\frac{dg_s}{dT}\right)dT\bigg],
\label{eq:HstH0}
\end{align}
where, $\Omega_{r0}\simeq 8.5\times10^{-5}$ is the current value of fractional energy density of radiation and $T_r = 10^4$ GeV has been used in this calculation. Collating these results, the GW spectrum observed today is given by,
\begin{align}
\Omega_{\rm gw} = \Omega_{r0} \Omega_{\rm gw*} & \exp\bigg[\int_{T_*}^{T_r}
  \frac{4}{T'}\left(1+ \frac{T}{3g_s}\frac{dg_s}{dT}\right)\, dT\bigg] \nn \\
&\times \exp\bigg[\int_{T_{r}}^{T_{*}}
  \frac{3}{T}\,(1+w_{\rm eff})
  \left(1+ \frac{T}{3g_s}\frac{dg_s}{dT}\right)\, dT\bigg] .
\label{eq:gw-spec}
\end{align}
Few comments at this point are in order: recent lattice results have consistently reported a non-zero trace anomaly and in fact, it is large near transition temperature. However, trace anomaly vanishes at $T\gg T_c$. Thus, the effect of trace anomaly will be important near QCD transition and must be taken into account in all realistic calculations as considered in \cite{Drees:2015exa} for dark matter relic density calculation. As mentioned earlier, the ratio of the scale factor at the epoch of transition to that of its value today is given as, $a_*/a_0 = \exp[{\int_{T_*}^{T_0} \frac{1}{T}\big(1+ \frac{T}{3g_s}\frac{dg_s}{dT}\big)\, dT}]$ $= (g_s(T_0)/g_s(T_*))^{1/3} T_0/T_*$ and does not change appreciably. As a result, the peak frequency of the gravitational wave redshifted to today's epoch is 
 \be
 \nu_{\rm peak} = \nu_*\, \bigg(\frac{a_*}{a_0}\bigg)=\nu_*\,
 \exp\left[{\int_{T_*}^{T_0} \frac{1}{T}\big(1+ \frac{T}{3g_s}\frac{dg_s}{dT}\big)\, dT}\right].
 \ee
However, the frequency of the gravitational radiation  generated at the epoch of  transition $\nu_*$ is set by the Hubble parameter at the epoch of phase transition. Hence, change in $H_*$ will  change the frequency of the gravitational radiation.
Presence of trace anomaly will alter the fractional energy density of the GW, $\Omega_{\rm gw} = \Omega_{\rm gw*}\, (a_*/a_0)^4\, (H_*/H_0)^2$. In fact, the fractional energy density of the GW will be enhanced. This enhancement is due to the fact that the effective equation of state parameter $w_{\rm eff}$, which depicts the energy content of the universe and hence governs the background evolution, decreases from $1/3$, the ideal value. This implies that the density will fall slower than $a^{-4}$ and thus, the Hubble parameter will change slower than $T^2$ (see Fig.~(\ref{fig:ast-a0})) which implies that the value of Hubble parameter at the time of transition $H_*$ will be higher than its value obtained without QCD equation of state. Consequently, there will be an overall enhancement in the GW signal.

\section{Sources of gravitational wave} 
\label{s:sourcegw}
With this, we now turn our attention to $\Omega_{\rm gw}(T_*)\equiv \Omega_{\rm gw*}$, which is important for the determination of the GW spectrum. Various sources have already been reported in the literature that contribute to $\Omega_{\rm gw*}$. To name a few, cosmic strings and domain walls \cite{Vachaspati:1984gt, Vachaspati:1984yi}, solitons and soliton stars \cite{Gleiser:1989mb} and cosmological phase transitions \cite{1986MNRAS.218..629H}. However, we consider the stochastic background arising from the first order phase transition. Further, we consider two important processes involved in the production of SGW at first order phase transition, namely (i) the collision of bubble walls and shocks in the plasma \cite{1986MNRAS.218..629H, Kosowsky:1992rz, Kosowsky:1991ua, Kamionkowski:1993fg, Caprini:2007xq, Huber:2008hg}
and (ii) Magnetohydrodynamic (MHD) turbulence in the plasma after bubble collision~\cite{Caprini:2006jb}. We refer to \cite{Caprini:2015zlo}, and references therein, for detailed discussion.

\begin{figure}[!t]
\hspace{-0.5cm}
\subfloat[Subfigure 1 list of figures text][]{
\includegraphics[width=3.2in,height=2.6in,angle=0]{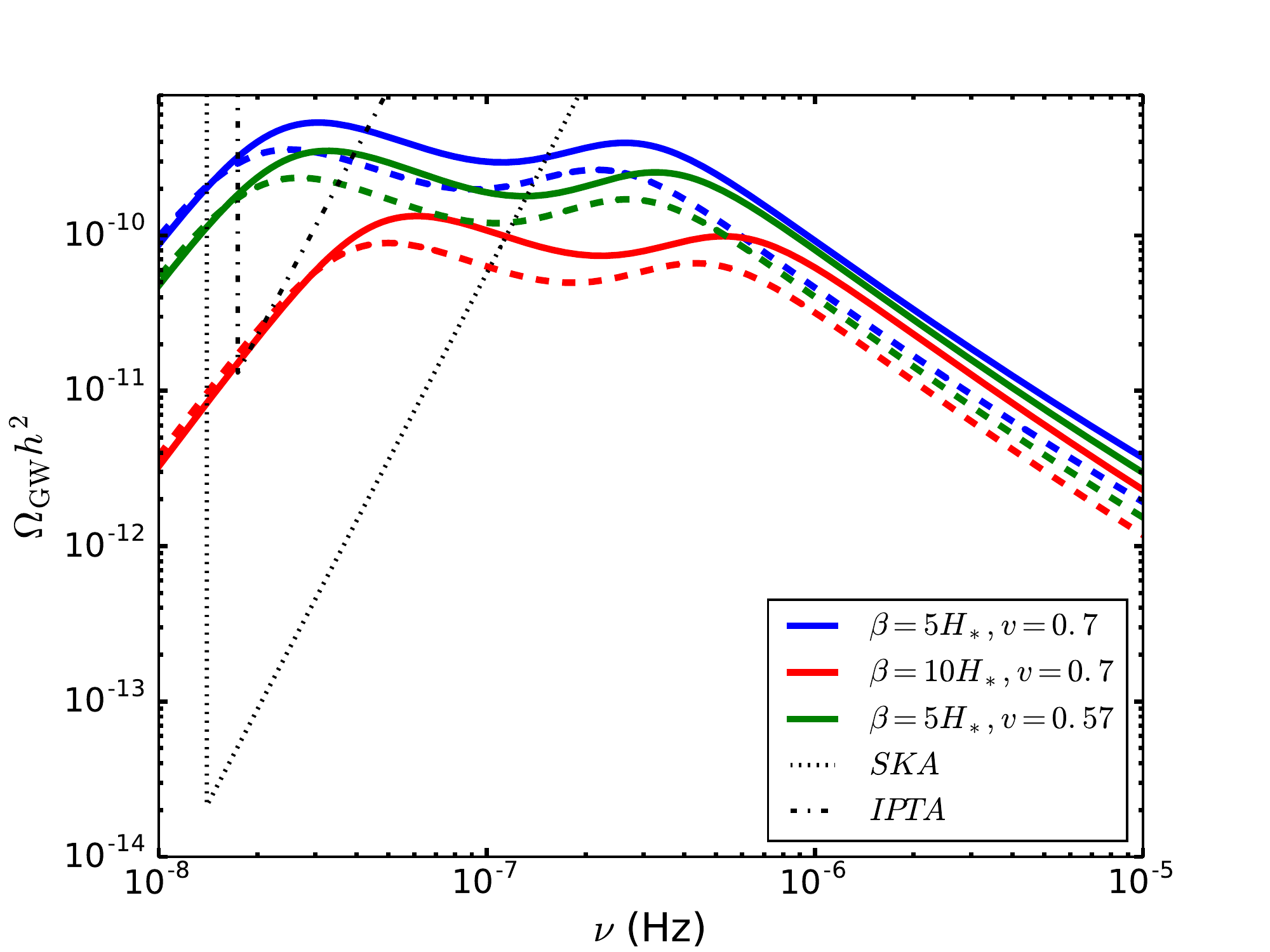}
\label{fig:gw-spec}
}~~~~
\hspace{-0.6cm}
\subfloat[Subfigure 2 list of figures text][]{
\includegraphics[width=3.2in,height=2.6in,angle=0]{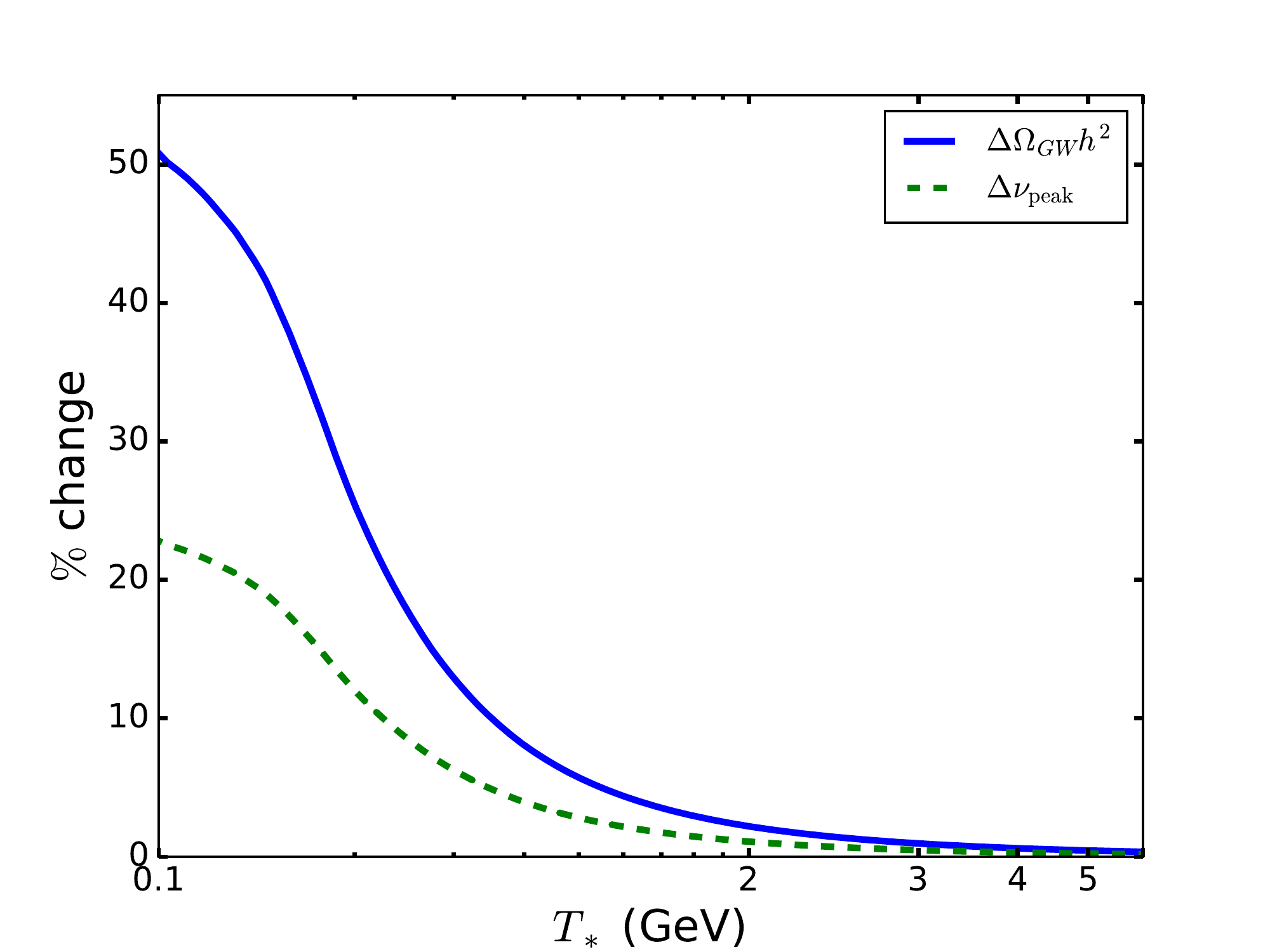}
\label{fig:per-chng}
	  }
\caption{(a) Net contribution to the GW spectrum due bubble collision and MHD turbulence. Solid lines denotes the net contribution to the GW signal with trace anomaly and corresponding dashed line denotes the same without trace anomaly. The projected reach of IPTA~\cite{IPTA-2013} and  SKA \cite{Dewdney-ska, Kramer:2004hd} GW detection experiments  are respectively indicated by dotted and dot dashed lines; (b) percentage change in the GW signal (blue solid) and in the peak frequency redshifted to today's epoch $\nu_{\rm peak}$ (green dashed).}
\end{figure}

\subsection{Bubble collisions} 
\label{sb:bubble}
Contribution to the gravitational waves spectrum from bubble  collisions, under envelope approximation, can be well described by~\cite{Huber:2008hg}:
  \be
  \Omega_{\rm gw*}^{(b)}(\nu) =\bigg(\frac{H_*}{\beta}\bigg)^2
  \bigg(\frac{ \kappa_b\alpha}{1+\alpha}\bigg)^2
  \bigg(\frac{0.11 v^3}{0.42 + v^2}\bigg)
 \, S_b(\nu),
  \ee
  where $\beta^{-1}$ is the time duration of the phase transition, $\alpha$ is the ratio of the vacuum energy density released in the phase transition to that of the radiation, $v$ is the wall velocity and $\kappa_b$ denotes the fraction of the latent heat of the phase transition deposited on the bubble wall. The function $S_b(\nu)$ parametrizes the spectral shape of the SGW background which is, obtained by fitting simulation data ~\cite{Huber:2008hg} as well as analytically \cite{Jinno:2016vai}, given as
  \be
  S_b(\nu) =\frac{3.8\,(\nu/\nu_b)^{2.8}}{1 + 2.8\,(\nu/\nu_b)^{3.8}}\, ,
  \ee
  where $\nu_b = \nu_{b*} (a_*/a_0)$ with $\nu_{b*}$ being the peak frequency of the contribution to the SGW spectrum from bubble collisions at the time of phase transition. An approximate form of $\nu_{b*}$ is given as \cite{Huber:2008hg},
\be
  \frac{\nu_{b*}}{\beta} = \frac{0.62}{(1.8-0.1 v + v^2)}.
\ee

\subsection{Magnetohydrodynamical (MHD) turbulence} 
\label{sb:mhd}
During QCD phase transition, the kinetic and magnetic Reynolds number of cosmic fluid are large~\cite{Caprini:2009yp}. Therefore, the percolation of the bubbles can generate magnetohydrodynamical (MHD) turbulence in the fully ionized plasma. Assuming Kolmogorov-type turbulence as proposed in \cite{Kosowsky:2001xp}, the contribution to the gravitational wave spectrum is described by~\cite{Caprini:2009yp,Binetruy:2012ze}:
 \be
  \Omega_{\rm gw*}^{({\rm mhd})}(\nu) = \,\bigg(\frac{H_*}{\beta}\bigg)\,
  \bigg(\frac{\kappa_{{\rm mhd}}\,\alpha}{1+\alpha}\bigg)^{3/2}\, v\,
  S_{{\rm mhd}}(\nu),
  \ee
  where $\kappa_{{\rm mhd}}$ is the fraction of latent heat converted into the turbulence. Shape of the spectra is determined by
  
\be
S_{{\rm mhd}}(\nu) =
\frac{(\nu/\nu_{{\rm mhd}})^3}{[(1 +\nu/\nu_{{\rm mhd}})]^{11/3}\, (1 + 8\pi\,\nu/{\cal H}_*)}\, ,
\ee

where ${\cal H}_* = (a_*/a_0) H_*$~\cite{Caprini:2009yp, Binetruy:2012ze}. The peak frequency redshifted to today's epoch is given as $\nu_{{\rm mhd}} =\nu_{{\rm mhd}*}(a_*/a_0)$, with $\nu_{{\rm mhd}*}/\beta = (3.5/2)\,v^{-1}$.
Features of the phase transition dynamics {\it e.g.}, wall velocity $v$, duration of phase transition $\beta^{-1}$, efficiency $\kappa$ and $\alpha$ play very crucial role in determining the peak position and amplitude of the GW signal.  However, there is no  robust way to determine $\kappa$ as yet. Therefore, we simplify the calculations by assuming that  $(\kappa_b\alpha)/(1+\alpha) =  (\kappa_{\rm mhd}\alpha)/(1+\alpha)$. Paucity of precise model for QCD phase transition makes it difficult to choose the duration of phase transition as well. Although, \cite{Hogan:1984hx} argues in favor of large $\beta/H_*$ but smaller values of $\beta/H_*$ can not be simply ruled out. We take two sets of values, $\beta = 5 H_*$ and $\beta = 10H_*$ in this calculation.

\section{Results} 
\label{s:result}
In Fig.~(\ref{fig:gw-spec}), we show the net GW spectrum due to bubble collision and MHD turbulence for different values of $\beta$ and $v$ with $(\kappa_b\alpha)/(1+\alpha) =  0.05$. Solid lines are the new results with QCD equation of state while corresponding dashed lines are the results with the ideal gas equation of state where $p=\rho/3$. Note that smaller value of $\beta/H_*$ implies larger duration of transition and hence larger amplitude of the signal. Similarly, larger wall velocity will produce larger amplitude. Also, the enhancement of the signal is due to the change in the background evolution. As the frequency of the GW signal depends on the Hubble parameter at the epoch of transition, we found that the peak frequency redshifted to todays epoch changes by $\sim 25\%$ and the  signal enhances by $\sim 50\%$ (see Fig.~(\ref{fig:per-chng})).

\section{Summary and conclusions} 
\label{s:concl}
To conclude, recent lattice calculations show the presence of trace anomaly near QCD phase transition. We have revisited the SGW power spectrum, generated by strong first order phase transition around QCD epoch, in presence of trace anomaly. We have shown that the trace anomaly leads to a slow down in the expansion of the universe which in turn enhance the gravitational wave signal (see Fig.~(\ref{fig:gw-spec})). This enhancement is around $\sim 50\%$ (see Fig.~(\ref{fig:per-chng})) which is significantly large. Characteristic frequency of the signal observed today is set by the Hubble parameter at the epoch of transition. Therefore, the frequency redshifted to current epoch also changes by $\sim 25\%$. Thus, the possibility of detection of the stochastic gravitational background signal due to phase transition in future observations is more. This result will be important while comparing observations from Pulsar Timing Array ~\cite{IPTA-2013} and Square Kilometer Array  \cite{Dewdney-ska, Kramer:2004hd} with the predictions of phase transitions.
\acknowledgments
The work of UKD is supported by Department of Science and Technology, Government of India under the fellowship reference number PDF/2016/001087 (SERB National Post-Doctoral fellowship).

\bibliographystyle{JHEP} 
\bibliography{ref} 

\end{document}